\documentclass[%
  twocolumn,
 showpacs,
 showkeys,
 preprintnumbers,
 amsmath,amssymb,
 aps,
 pra,
  longbibliography,
 ]{revtex4-1}

\usepackage[dvipsnames]{xcolor}

\usepackage{mathptmx}

\usepackage{amssymb,amsthm,amsmath}

\usepackage{tikz}
\usetikzlibrary{calc,decorations.pathreplacing,decorations.markings,positioning,shapes}

\usepackage[breaklinks=true,colorlinks=true,anchorcolor=blue,citecolor=blue,filecolor=blue,menucolor=blue,pagecolor=blue,urlcolor=blue,linkcolor=blue]{hyperref}
\usepackage{graphicx}
\usepackage{url}

\colorlet{BLUE}{blue}

\newcommand{\XX}{\hphantom{-}} 

\begin{document}

\title{Completing bases in four dimensions}

\author{Hans Havlicek}
\email{havlicek@geometrie.tuwien.ac.at}
\affiliation{Institute of Discrete Mathematics and Geometry,
TU Wien, Wiedner Hauptstrasse 8-10/104,
1040 Vienna, Austria}
\homepage{https://www.geometrie.tuwien.ac.at/havlicek/}

\author{Karl Svozil}
\email{svozil@tuwien.ac.at}
\affiliation{Institute for Theoretical Physics,
TU Wien,
Wiedner Hauptstrasse 8-10/136,
1040 Vienna,  Austria}
\homepage{http://tph.tuwien.ac.at/~svozil}

\date{\today}

\begin{abstract}
Criteria and constructive methods for the completion of an incomplete basis of, or context in, a four-dimensional Hilbert space by (in)decomposable vectors are given.
\end{abstract}

\keywords{Greenberger-Horne-Zeilinger argument, Gleason theorem, Kochen-Specker theorem, Born rule, gadget graphs, Greechie diagram, McKay-Megill-Pavicic diagram (MMP), orthogonality hypergraph}
\pacs{03.65.Ca, 02.50.-r, 02.10.-v, 03.65.Aa, 03.67.Ac, 03.65.Ud}

\maketitle

\section{Completion of incomplete contexts}

We shall find and analyze orthogonal vectors spanning two-dimensional subspaces
of four-dimensional real or complex Hilbert space
that are orthogonal to a given two-dimensional subspace.
In particular, we are interested whether those vectors are
indecomposable---a property of pure state vectors signifying entanglement of multipartite quantized systems.

In physics, this question is pertinent to a variety of tasks:
First, any orthonormal basis can be, by dyadic or tensor products, rewritten as a system of mutually perpendicular orthogonal projection operators.
This system can be extended to maximal hermitian operators associated with maximal quantum observables
in terms of their spectral sums containing mutually distinct eigenvalues.
Often these maximal operators are denoted as, and identified with, quantum mechanical contexts.

Quantum contexts serve as the basic building blocks of quantum logical and probabilistic certification of quantization.
Boole-Bell type arguments consider three or more isolated contexts and compare classical with quantum predictions of expectation functions.
Hardy-type arguments involve multiple intertwining contexts with two endpoints, such that classical predictions relate the truth values of these endpoints.
Intertwining contexts with ``scarce'' two-valued states---featuring classical nonseparability of elementary propositions
or nonunital sets of two-valued states interpretable as classical (truth) value assignments---yield
logics that cannot be homomorphically embedded into ``larger'' Boolean algebras.
And Kochen-Specker-type arguments demonstrate the total absence of any classical interpretation in terms of the aforementioned two-valued states.

All of the above tactics to certify quantization need quantum representations of contexts in terms of (intertwining) orthonormal bases.
Often the algebraic structures are formulated and depicted in terms of (hyper)graphs~\cite{Bretto-MR3077516}.
These hypergraphs, to be realizable in terms of quantum observables,
need to allow a faithful orthogonal representation~\cite{lovasz-79,GroetschelLovaszSchrijver1986}, essentially a vertex labeling
by vectors, such that adjacent vertices correspond to orthogonal vectors.
Although in principle the equations resulting from such relations may be solvable, their direct solution turns out to be unattainable.
Therefore one is left with heuristic methods of parametrization~\cite{Pavii2018} that yield incomplete orthonormal systems;
and therefore the necessity to complete those findings by supplementing missing base vectors.

In four dimensions, concerning indecomposability---or, in physical terms, entanglement---this task is straightforward
for three given mutually orthogonal unit vectors---the one-dimensional subspace
spanned by the missing vector is uniquely defined, and there is no choice.
However, a completion with (in)decomposable vectors is not straightforward for two given unit vectors.
As we shall see there are rather subtle criteria of
(in)decomposability if the four-dimensional Hilbert space is interpreted as a
tensor product of two two-dimensional spaces.

Such analysis is pertinent to the aforementioned task of
completing one or more bases or contexts of a (hyper)graph: find a complete
faithful orthogonal representation (aka coordinatization) of a hypergraph when
only a coordinatization of the intertwining observables is known.
For instance, for Hardy type arguments, it is significant
whether the resulting completion of the context may comprise
(in)decomposable vectors~\cite{svozil-2020-hardy}.

We shall, in particular, consider a four-dimensional real or complex Hilbert
space $\mathcal{H}$, where $\mathcal{H}$ is either the column space $\mathbb
R^4$ or $\mathbb C^4$. Suppose further that two unit vectors ${\bf e}_1$ and
${\bf e}_2$ are known which are orthogonal, such that $\langle {\bf e}_1 \vert
{\bf e}_2 \rangle =0$. An orthogonal basis can be formed with these two known
vectors, as well as with two ``missing'' vectors ${\bf a}$ and ${\bf b}$. Those
latter missing vectors ought to have additional properties we are interested
in; in particular, for Hilbert spaces which can be considered as tensor
products of smaller-dimensional spaces, (in)decomposability.

One uniform way of finding the general form of the missing vectors is by
arranging a to-be-completed orthonormal basis (aka context) $B=\big\{ {\bf
e}_1,{\bf e}_2,{\bf a},{\bf b} \big\}$ into a unitary matrix $
\textsf{\textbf{U}} =
\begin{pmatrix}
{\bf e}_1,{\bf e}_2,{\bf a},{\bf b}
\end{pmatrix}
$ and solving~\cite[Theorem~2]{Alkan2017}
\begin{equation}
\left\vert \textrm{det} \left( \textsf{\textbf{U}}  \right) \right\vert =
\frac{1}{4} \textrm{Tr}\left( \textsf{\textbf{U}}\textsf{\textbf{U}}^\dagger  \right)=1,
\label{2020-e-unitary}
\end{equation}
where ``$\dagger$'' stands for transposition and complex conjugation (which, in
the real case, reduces to transposition ``$^\intercal$'').

The subspace $\mathcal{M}^\perp$ of the Hilbert space $\mathcal{H}$ orthogonal
to the subspace $\mathcal{M}$ spanned by both ${\bf e}_1$ and ${\bf e}_2$ will
be two-dimensional and spanned by ${\bf a}$ and ${\bf b}$. This leaves a
continuity of freedom in choosing those latter vectors.

Continuity of choices aside; whether or not the missing vectors can be selected
to be (in)decomposable is not merely a question of choice but depends on the
vectors ${\bf e}_1$ and ${\bf e}_2$ one started with. For instance, if $ {\bf
e}_1=
\begin{pmatrix}
1,0,0,0
\end{pmatrix}^\intercal
$
and
$
{\bf e}_2=
\begin{pmatrix}
0,1,0,0
\end{pmatrix}^\intercal
$
an elementary calculation shows that there is no option for both ${\bf a}$ as
well as ${\bf b}$ not to be decomposable: Suppose $ {\bf a} \propto
\begin{pmatrix}
a_{1},a_{2},a_{3},a_{4}
\end{pmatrix}^\intercal
$; then $ \langle {\bf a} \vert {\bf e}_1 \rangle = \langle {\bf a} \vert {\bf
e}_2 \rangle = 0 $ implies $a_{1} = a_{2} =0$. Therefore, ${\bf a}$ must be of
the form $\begin{pmatrix} 0,0,a_{3},a_{4}
\end{pmatrix}^\intercal $.
The same argument holds for ${\bf b}$. By the criterion of decomposability for
vectors derived later, by which the scalar product of the two ``outer''
components of the vector must be equal to the two ``inner'' components of the
vector, both ${\bf a}$ as well as ${\bf b}$ are decomposable.

For the sake of an example in which ${\bf a}$ as well as ${\bf b}$
may either be chosen decomposable or indecomposable, consider
the instance
$
{\bf e}_1=
\begin{pmatrix}
1,0,0,0
\end{pmatrix}^\intercal
$
and
$
{\bf e}_2=
\begin{pmatrix}
0,0,0,1
\end{pmatrix}^\intercal
$ which allows either
decomposable completions such as
$
{\bf a}=
\begin{pmatrix}
0,1,0,0
\end{pmatrix}^\intercal
$
and
$
{\bf b}=
\begin{pmatrix}
0,0,1,0
\end{pmatrix}^\intercal
$
or indecomposable completions such as
$
{\bf a}=\frac{1}{\sqrt{2}}
\begin{pmatrix}
0,1,1,0
\end{pmatrix}^\intercal
$
and
$
{\bf b}= \frac{1}{\sqrt{2}}
\begin{pmatrix}
0,1,-1,0
\end{pmatrix}^\intercal
$.

The general question therefore remains: given two orthogonal unit vectors ${\bf
e}_1$ and ${\bf e}_2$, when is it possible for those missing vectors ${\bf a}$
and ${\bf b}$ of a ``completed'' orthogonal basis (aka context) $B=\big\{ {\bf
e}_1,{\bf e}_2,{\bf a},{\bf b} \big\}$ to be (in)decomposable?

\section{Nomenclature}\label{se:II}

Let $\mathcal{H}_2$ and $\mathcal{H}$ either denote the real vector spaces
$\mathbb{R}^2$ and $\mathbb{R}^4$ or the complex vector spaces $\mathbb{C}^2$
and $\mathbb{C}^4$. The standard inner product $\langle\cdot\vert\cdot\rangle$
makes $\mathcal{H}_2$ and $\mathcal{H}$ into a Hilbert space.
We identify the outer or
tensor product $\mathcal{H}_2\otimes\mathcal{H}_2$ with $\mathcal{H}$ as
follows. Given vectors $ {\bf u} =
\begin{pmatrix}
u_{1},u_{2}
\end{pmatrix}^\intercal
$ and $ {\bf v} =
\begin{pmatrix}
v_{1},v_{2}
\end{pmatrix}^\intercal
$ in $\mathcal{H}$ we let ${\bf u} \otimes {\bf v} =
\begin{pmatrix}
u_{1}v_{1},u_{1}v_{2},u_{2}v_{1},u_{2}v_{2}
\end{pmatrix}^\intercal \in\mathcal{H}
$, which is a form of ``vectorization'' (that is, a flattening) of this tensor
product.
This product can be compared to the general form of a vector in four dimensions
${\bf z} =
\begin{pmatrix}
z_{1},z_{2},z_{3},z_{4}
\end{pmatrix}^\intercal
$.
Therefore, for ${\bf z}$ to be decomposable
$z_{1}= x_{1}y_{1}$,
$z_{2}= x_{1}y_{2}$,
$z_{3}= x_{2}y_{1}$, and
$z_{4}= x_{2}y_{2}$,
from which, because of commutativity of scalar multiplication, follows that
\begin{equation}
z_{1}z_{4}= x_{1}y_{1} x_{2}y_{2} = x_{1}x_{2}y_{1}y_{2}= x_{1}y_{2}x_{2}y_{1} = z_{2}z_{3}.
\label{2020-dex-e-di4d}
\end{equation}
That is, the product of the ``outer components'' $z_{1}z_{4}$ of ${\bf z}$ must
be equal to the product of its ``inner components'' $z_{2}z_{3}$, or
equivalently, $z_{1}z_{4}-z_{2}z_{3}=0$~\cite[p.~18]{mermin-07}.
This condition is also sufficient, as
it renders three equations for the four unknowns $x_{1}$, $x_{2}$,
$y_{1}$ and $y_{2}$.

Criterion~(\ref{2020-dex-e-di4d}) for decomposability
can be rewritten in terms of a symmetric bilinear form as follows.
The mapping ${\bf z}=
\begin{pmatrix}
z_{1},z_{2},z_{3},z_{4}
\end{pmatrix}^\intercal  \mapsto 2(z_{1}z_{4}-z_{2}z_{3})$
is a quadratic form which has an associated bilinear form (not to be confused
with the scalar or inner product denoted by $\langle{\bf a}\vert {\bf b}\rangle$)
\begin{equation}
\begin{aligned}
&({\bf a} \vert {\bf b})
 =
\left(
a_1b_4  - a_2b_3 - a_3b_2 + a_4b_1
\right) \\
&\qquad =
\begin{pmatrix}
a_{1},a_{2},a_{3},a_{4}
\end{pmatrix}
\begin{pmatrix}
0&\XX 0&\XX 0&\XX 1\\
0&\XX 0&-1&\XX 0\\
0&-1&\XX 0&\XX 0\\
1&\XX 0&\XX 0&\XX 0\\
\end{pmatrix}
\begin{pmatrix}
b_{1}\\b_{2}\\b_{3}\\b_{4}
\end{pmatrix} \\
& \qquad ={\bf a}^\intercal  \cdot \textsf{\textbf{A}} \cdot{\bf b}
, \\
&\textrm{with }
\textsf{\textbf{A}} :=
\begin{pmatrix}
0&\XX 0&\XX 0&\XX 1\\
0&\XX 0&-1&\XX 0\\
0&-1&\XX 0&\XX 0\\
1&\XX 0&\XX 0&\XX 0\\
\end{pmatrix}
.
\end{aligned}
\label{2020-dex-e-di4dblf}
\end{equation}
Therefore,
\begin{equation}
( {\bf z} \vert {\bf z} )=  2 ( z_1z_4  - z_2z_3  ) = 0
\label{2020-dex-e-di4dblfnccf}
\end{equation}
characterises ${\bf z}$ as decomposable.

The (non-degenerate) bilinear form~(\ref{2020-dex-e-di4dblf}) can then be used
to define a Gramian matrix of two vectors ${\bf a}$ and ${\bf b}$ by
\begin{equation}
G_{{\bf a}{\bf b}}
=
\begin{pmatrix}
({\bf a} \vert {\bf a})  & ({\bf a} \vert {\bf b})\\
({\bf b} \vert {\bf a})  & ({\bf b} \vert {\bf b})
\end{pmatrix}
.
\label{2020-dex-e-gramian}
\end{equation}
This definition of the Gramian matrix for two vectors has a straightforward
generalization for an arbitrary finite number of vectors which we shall use
later.

Because of symmetry $({\bf a} \vert {\bf b})= ({\bf b} \vert {\bf a})$ the Gram determinant satisfies
\begin{equation}
\begin{vmatrix}
({\bf a} \vert {\bf a})  & ({\bf a} \vert {\bf b})\\
({\bf b} \vert {\bf a})  & ({\bf b} \vert {\bf b})
\end{vmatrix}
=
({\bf a} \vert {\bf a})({\bf b} \vert {\bf b})  - ({\bf a} \vert {\bf b})^2
.
\label{2020-dex-e-gramian-determinant}
\end{equation}

The symmetric matrix $\textsf{\textbf{A}}$ that is defined
in~(\ref{2020-dex-e-di4dblf}) coincides with its inverse
$\textsf{\textbf{A}}^{-1}$.
Let $\overline{x}= \Re x - i \Im x$ stand for
complex conjugation, so that, for real vector spaces, $\overline {\bf x} = {\bf
x}$ for all vectors {\bf x}. The matrix $\textsf{\textbf{A}}^{-1}$
defines a bijection
\begin{equation}
\begin{aligned}
{\bf x}
=
\begin{pmatrix}x_1,x_2,x_3,x_4\end{pmatrix}^\intercal
\mapsto &&
\textsf{\textbf{A}}^{-1} \cdot
\overline{{\bf x}}
=
\begin{pmatrix}\overline{x}_4,-\overline{x}_3,-\overline{x}_2,\overline{x}_1\end{pmatrix}^\intercal
\\
\; &&=
\begin{pmatrix}{x_4},-{x_3},-{x_2},{x_1}\end{pmatrix}^\dagger
=:
\widetilde {\bf x}
,
\end{aligned}
\label{2020-dec-def-x-tilde}
\end{equation}
which is linear in the real case and antilinear in the complex case.

For two vectors ${\bf x}$ and ${\bf y}$, because of $\textsf{\textbf{A}}^{-1}
=\left(\textsf{\textbf{A}}^{-1}\right)^\intercal $,
\begin{equation}
\begin{aligned}
    \langle  {\bf x}\vert {\bf y} \rangle
&=
     {\bf x}^\dagger\cdot {\bf y} =
     {\bf x}^\dagger \cdot \left(\textsf{\textbf{A}}^{-1} \cdot \textsf{\textbf{A}}\right) \cdot {\bf y} \\
&=
    \overline{{\bf x}}^\intercal  \cdot \left(\textsf{\textbf{A}}^{-1}\right)^\intercal \cdot \textsf{\textbf{A}}\cdot {\bf y} \\
&=
    \left[\left(\textsf{\textbf{A}}^{-1}\right)\cdot \overline{{\bf x}}\right]^\intercal \cdot  \textsf{\textbf{A}}  \cdot {\bf y} =
    (\widetilde {\bf x}\vert {\bf y}),
\end{aligned}
\label{2020-dec-e-transcript}
\end{equation}
where $\widetilde {\bf x} = \left(\textsf{\textbf{A}}^{-1}\right)\cdot \overline{{\bf x}}$.
That is, the inner product can be rewritten in terms of the
bilinear form~(\ref{2020-dex-e-di4dblf})
which enters the Gram matrix~(\ref{2020-dex-e-gramian}).
This is a central facility for the following classification
of two-dimensional planes in four-dimensional Hilbert space.

As a side note observe that, though a coordinate change in the new coordinates
${\bf x} \mapsto {\bf x}'$ with $ x_1 \mapsto x_1' = (x_1+x_4)/\sqrt{2} $, $
x_2 \mapsto x_2' = (x_2-x_3)/\sqrt{2} $, $ x_3 \mapsto x_3' =
(x_2+x_3)/\sqrt{2} $, and $ x_4 \mapsto x_4' = (x_1-x_4)/\sqrt{2} $ ``mixing
outer as well as inner components, respectively'', this symmetric bilinear form
can be rewritten in terms of a diagonal matrix $ \textsf{\textbf{A}}'=
\textrm{diag} (1,1,-1,-1) $, such that $( {\bf x} \vert {\bf y} ) = \left({\bf
x}'\right)^\intercal \cdot \textsf{\textbf{A}}' \cdot {\bf y}'$; and, in
particular, $( {\bf z} \vert {\bf z}) = \left({\bf z}'\right)^\intercal \cdot
\textsf{\textbf{A}}' \cdot {\bf z}'$. {\large[}For a proof, expand $\left({\bf
x}'\right)^\intercal \cdot \textsf{\textbf{A}}' \cdot {\bf y}'$ in terms of
${\bf x}$ and ${\bf y}$.{\large]}
In these new coordinates ${\bf z}'$ a necessary and sufficient criterion for
${\bf z}$ to be decomposable is $\left({\bf z}'\right)^\intercal \cdot
\textsf{\textbf{A}}' \cdot {\bf z}' = 0$.

Over the complex numbers only, a second coordinate change
 ${\bf x}' \mapsto {\bf x}''$
with $ x_1' \mapsto x_1'' = x_1'$, $ x_2' \mapsto x_2'' = x_2'$, $ x_3' \mapsto
x_3'' = i x_3'$, and $ x_4' \mapsto x_4'' = i x_4''$ yields
$\textsf{\textbf{A}}''=\textrm{diag} ( 1,1,1,1 )$ as matrix of this symmetric
bilinear form.

We shall make use of the following equality. Let $ {\bf s} =
\begin{pmatrix} s_{1},s_{2}
\end{pmatrix}^\intercal
$, $ {\bf t} =
\begin{pmatrix}
t_{1},t_{2}
\end{pmatrix}^\intercal
$, $ {\bf u} =
\begin{pmatrix}
u_{1},u_{2}
\end{pmatrix}^\intercal
$ and $ {\bf v} =
\begin{pmatrix}
v_{1},v_{2}
\end{pmatrix}^\intercal
$ be arbitrary vectors of $\mathcal{H}_2$. Then $\sum_{j,k=1}^2
\overline{(s_jt_k)}(u_jv_k) = \Big(\sum_{j=1}^2 \overline{s}_j u_j\Big) \Big(
\sum_{k=1}^2\overline{t}_k v_k \Big)$ implies
\begin{equation}\label{hans-inner-rel}
    \langle {\bf s}\otimes {\bf t}\vert{\bf u}\otimes{\bf v} \rangle
    = \langle {\bf s}\vert{\bf u}\rangle \langle {\bf t}\vert{\bf v} \rangle .
\end{equation}
Eq.~\eqref{hans-inner-rel} can be rephrased in the following way. The inner
product on $\mathcal{H}$ is the second tensor power of the inner product on
$\mathcal{H}_2$; see \cite[Appendix~A, p.~164]{mermin-07} or \cite[Sect.~3.4,
pp.~47--48]{Weidmann1980}.

Likewise, $s_1 t_1 u_2 v_2 - s_1 t_2 u_ 2v_1 - s_2 t_1 u_1 v_2 + s_2 t_2 u_1
v_1 = (s_1 u_2 - s_2 u_1)(t_1 v_2 - t_2 v_1)$ results in
\begin{equation}\label{hans-bilinear-det}
    ( {\bf s}\otimes {\bf t}\vert{\bf u}\otimes{\bf v} )
    = \det( {\bf s},{\bf u})\det({\bf t},{\bf v} ) ,
\end{equation}
where $({\bf s},{\bf u})$ stands for the matrix whose first and second column
are ${\bf s}$ and ${\bf u}$, respectively. That is, the symmetric bilinear form
$(\cdot\vert\cdot)$ on $\mathcal{H}$ from \eqref{2020-dex-e-di4dblf} is the
second tensor power of the skew-symmetric bilinear form given by the
determinant on $\mathcal{H}_2$~\cite[Sect.~1.22, p.~30--31]{book:4317}.

The (anti)linear transformation of $\mathcal{H}_2$ sending $ {\bf u} =
\begin{pmatrix}
    u_{1},u_{2}
\end{pmatrix}^\intercal
$ to $ {\bf u}^\times =
\begin{pmatrix}
    \overline{u}_{2},-\overline{u}_{1}
\end{pmatrix}^\intercal
$ satisfies
\begin{equation}\label{hans-orthogonal}
\langle {\bf u}\vert {\bf u}^\times\rangle = 0 .
\end{equation}
Furthermore, it allows us to rewrite any inner product in terms of the
determinant:
\begin{equation}\label{hans-inner-det}
    \langle {\bf u}\vert {\bf v}\rangle =
\overline{u}_{1}v_1 + \overline{u}_{2}v_2 =
    \begin{vmatrix}
     \XX\overline{u}_{2}  & {v}_{1}\\
     -\overline{u}_{1}  & {v}_{2}
    \end{vmatrix}
    = \det({\bf u}^\times, {\bf v}) .
\end{equation}
We also observe that
\begin{equation}\label{hans-inner-x}
    \langle {\bf u}^\times\vert {\bf v}^\times\rangle =
    u_2\overline{v}_2 + u_1\overline{v}_1 =
    \overline{\langle {\bf u}\vert {\bf v}\rangle} =
    \langle {\bf v}\vert {\bf u}\rangle .
\end{equation}
Furthermore, the second tensor power of the (anti)linear transformation
${\bf u}\mapsto {\bf u}^\times$ equals the (anti)linear transformation from
Eq.~\eqref{2020-dec-def-x-tilde}:
\begin{equation}\label{hans-tilde-times}
    {\bf u}^\times \otimes {\bf v}^\times =
    \begin{pmatrix}
    \overline{u}_{2}\overline{v}_{2}, -\overline{u}_{2}\overline{v}_{1}, -\overline{u}_{1}\overline{v}_{2},
    \overline{u}_{1}\overline{v}_{1}
    \end{pmatrix}^\intercal
    = \widetilde{({\bf u} \otimes {\bf v})} .
\end{equation}

\section{Plane types}

Let $\mathcal{V}$ be any finite dimensional vector space over the real or
complex numbers. Basic results about symmetric bilinear forms on such a vector
space can be found, for example, in \cite[Theorems~11.21, 23, 24, 25, 26,
pp.~283--288]{roman-advancedLA}. We briefly recall these results in a form
which is tailored to our needs. That is, we consider a $k$-dimensional subspace
$\mathcal{S}$ of $\mathcal{H}$ together with the restriction of $(\cdot \vert
\cdot)$ to $\mathcal{S}$ rather than $\mathcal{V}$ together with an arbitrary
symmetric bilinear form on $\mathcal{V}$.

Suppose that an arbitrary basis of $\mathcal{S}$ is given. Then the Gramian
matrix of this basis with respect to $(\cdot \vert \cdot)$, which is defined in
analogy to~(\ref{2020-dex-e-gramian}), is a symmetric $(k\times k)$-matrix. In
a first step, one can switch to a (not necessarily orthogonal) basis $\{ {\bf
b}_1,{\bf b}_2,\ldots,{\bf b}_k \}$ of $\mathcal{S}$ which has a Gramian matrix
in diagonal form. Next, by scaling and reordering the vectors ${\bf b}_j$,
$j=1,2\ldots,k$, in an adequate way, one obtains a basis $\{ {\bf c}_1,{\bf
c}_2,\ldots,{\bf c}_k \}$ of $\mathcal{S}$ such that its Gramian matrix with
respect to $(\cdot \vert \cdot)$ takes the form
\begin{equation}\label{hans-real}
    \textrm{diag}(\underbrace{1,1,\ldots,1}_{p\geq 0},
                  \underbrace{-1,-1,\ldots,-1}_{r-p\geq 0},
                  \underbrace{0,0,\ldots,0}_{k-r\geq 0})
\end{equation}
for some $p,r$ in the real case and the form
\begin{equation}\label{hans-complex}
    \textrm{diag}(\underbrace{1,1,\ldots,1}_{r\geq 0},
                  \underbrace{0,0,\ldots,0}_{k-r\geq 0})
\end{equation}
for some $r$ in the complex case. (The need to distinguish between these two
cases stems from the fact that negative real numbers do not admit a real square
root.) The numbers $r$ and $p$ appearing in \eqref{hans-real} and
\eqref{hans-complex} are thereby uniquely determined by $\mathcal{S}$ and
$(\cdot \vert \cdot)$, that is, they do not depend on the choice of an
appropriate basis of $\mathcal{S}$. We also observe that the \emph{radical} of
$\mathcal S$, which is defined as
\begin{equation*}\label{}
    \mathrm{rad}(\mathcal{S}) =\{{\bf x}\in
\mathcal{S}
\mid ({\bf x} \vert {\bf y})=0 \mbox{~for all~}{\bf y}\in\mathcal{S} \} ,
\end{equation*}
satisfies
\begin{equation}\label{hans-radical=span}
    \mathrm{rad}(\mathcal{S}) = \mathrm{span}\{ {\bf c}_{r+1},{\bf c}_{r+2},\ldots,{\bf c}_k \}.
\end{equation}

In particular, letting $\mathcal{S}=\mathcal{H}$ yields the matrices
$\textsf{\textbf{A}}'=\textrm{diag}(1,1,-1,-1)$ (real case) and
$\textsf{\textbf{A}}''=\textrm{diag}(1,1,1,1)$ (complex case) that we already
encountered at the end of Section~\ref{se:II}.

Let $\mathcal{S}=\mathcal{M}$ be some plane, defined as a two-dimensional
subspace of the Hilbert space $\mathcal{H}$. Then, by the above, there exists
at least one basis $\{ {\bf c}_1,{\bf c}_2 \}$ of $\mathcal{M}$ such that the
Gramian matrix $G_{{\bf c}_1 {\bf c}_2}$ defined in~(\ref{2020-dex-e-gramian})
with respect to the bilinear form defined in~(\ref{2020-dex-e-di4dblf}) takes
on one of the following forms:
\begin{itemize}

\item[(i)]

real (Hilbert space) case:

\begin{itemize}

\item[(i.1)]  $G_{{\bf c}_1 {\bf c}_2} = \textrm{diag}(0,0)$,

\item[(i.2)] $G_{{\bf c}_1 {\bf c}_2} = \textrm{diag}(1,0)$,

\item[(i.3)] $G_{{\bf c}_1 {\bf c}_2} = \textrm{diag}(-1,0)$,

\item[(i.4)] $G_{{\bf c}_1 {\bf c}_2} = \textrm{diag}(1,1)$,

\item[(i.5)] $G_{{\bf c}_1 {\bf c}_2} = \textrm{diag}(-1,-1)$,

\item[(i.6)] $G_{{\bf c}_1 {\bf c}_2} = \textrm{diag}(1,-1)$;

\end{itemize}

\item[(ii)]

complex (Hilbert space) case:

\begin{itemize}

\item[(ii.1)] $G_{{\bf c}_1 {\bf c}_2} = \textrm{diag}(0,0)$,

\item[(ii.2)] $G_{{\bf c}_1 {\bf c}_2} = \textrm{diag}(1,0)$,

\item[(ii.3)] $G_{{\bf c}_1 {\bf c}_2} = \textrm{diag}(1,1)$.

\end{itemize}
\end{itemize}
If, say, the plane $\mathcal{M}$ is (uniquely) associated with the Gramian
matrix of the form $\textrm{diag}(1,0)$ then we shall denote $\mathcal{M}$ as a
{\em plane of type $(1,0)$}. The other cases are treated accordingly. Our
discussion below will establish the \emph{existence} of all these possible
plane types.

Let us come back to an earlier question: Suppose that an \emph{arbitrary} basis
$\{{\bf a},{\bf b}\}$ of $\mathcal{M}$ has been found. The question then is:
does this two-dimensional subspace allow or support (in)decomposable vectors in
four-dimensional space?

The cases (i.1)--(i.6) for real four-dimensional Hilbert space
$\mathcal{H}=\mathbb{R}^4$ as well as (ii.1)--(ii.3) for complex
four-dimensional Hilbert space $\mathcal{H}=\mathbb{C}^4$ discussed earlier
present a means to answer this question. Thereby the Gramian matrix $G_{{\bf a}
{\bf b}}$ is used for an identification and characterization of the particular
unique plane type of $\mathcal{M}$.

For real four-dimensional Hilbert space $\mathcal{H}=\mathbb{R}^4$ there are
six types of planes, corresponding to the cases~(i.1) to~(i.6) mentioned
earlier. In what follows three cases and the respective subcases will be
discussed which characterize those six plane types. We thereby apply results
that provide, for real vector spaces of any finite dimension, necessary and
sufficient conditions for the (semi)definiteness of a quadratic form in terms
of principal minors of its Gramian matrix with respect to an arbitrary basis.

In the following analysis, based on the earlier classification (i.1)--(i.6) for
real four-dimensional Hilbert space as well as (ii.1)--(ii.3) for complex
four-dimensional Hilbert space, the Gram determinant $\det (G_{{\bf a}{\bf
b}})$ will be denoted by $G$, and $G_{ij}$ stands for the element in the $i$th
row and the $j$th column of the Gramian matrix $G_{{\bf a}{\bf b}}$.

\subsection{Gram determinant $G > 0$, plane of types $(1,1)$ or $(-1,-1)$}

$G > 0$ means that $({\bf a} \vert {\bf a})$ as well as $({\bf b} \vert {\bf b})$ have
the same sign and are both non-zero.

\subsubsection{$G_{11} = ({\bf a} \vert {\bf a}) > 0$, plane of type~$(1,1)$}

In this subcase $({\bf a} \vert {\bf a})$ is positive, which indicates a
plane of type~$(1,1)$ \cite[Thm.~3, p.~306]{Gantmacher1}. Consequently, all non-zero
vectors of $\mathcal{M}$ are indecomposable.

A typical example is the two-dimensional subspace spanned by
${\bf a}=
\begin{pmatrix}
0,1,-1,0
\end{pmatrix}^\intercal
$
and
$
{\bf b}=
\begin{pmatrix}
1,0,0,1
\end{pmatrix}^\intercal
$. Any element of the span of ${\bf a}$ and
${\bf b}$ can be written as $
\begin{pmatrix}
x_1,x_2,-x_2,x_1
\end{pmatrix}^\intercal
$. The associated Gramian is of the form $ G_{{\bf a}{\bf b}}= \textrm{diag}
(2,2) $.

\subsubsection{$G_{11} = ({\bf a} \vert {\bf a}) < 0$, plane of type~$(-1,-1)$}

In this subcase $({\bf a} \vert {\bf a})$ is negative, which indicates a
plane of type~$(-1,-1)$ \cite[Thm.~5, p.~308]{Gantmacher1}. Consequently, all
non-zero vectors of $\mathcal{M}$ are indecomposable.

A typical example is the two-dimensional subspace spanned by
${\bf a}=
\begin{pmatrix}
0,1,1,0
\end{pmatrix}^\intercal
$
and
$
{\bf b}=
\begin{pmatrix}
1,0,0,-1
\end{pmatrix}^\intercal
$. Any element of the span of ${\bf a}$ and ${\bf b}$ can be written as $
\begin{pmatrix}
x_1,x_2,x_2,-x_1
\end{pmatrix}^\intercal
$.
The associated Gramian is of the form
$ G_{{\bf a}{\bf b}}= \textrm{diag} (-2,-2) $.

\subsection{Gram determinant $G = 0$,\\ plane of types $(0,0)$, $(1,0)$ or $(-1,0)$}

\subsubsection{$G_{11} = ({\bf a} \vert {\bf a}) = G_{22} = ({\bf b} \vert {\bf b})=0$, plane of type $(0,0)$}

In this subcase $G=({\bf a} \vert {\bf b})^2=0$, so that the Gramian vanishes --
that is,
$ G_{{\bf a}{\bf b}}= \textrm{diag} (0,0)$.
Hence,
by definition, $\mathcal{M}$ is a plane of type $(0,0)$. Any plane of this type
contains a continuity of decomposable vectors and no indecomposable vector.

A typical example is the two-dimensional subspace spanned by
${\bf a}=
\begin{pmatrix}
1,0,0,0
\end{pmatrix}^\intercal
$
and
$
{\bf b}=
\begin{pmatrix}
0,1,0,0
\end{pmatrix}^\intercal
$. Any element of the span of ${\bf a}$ and ${\bf b}$ can be written as $
\begin{pmatrix}
x_1,x_2,0,0
\end{pmatrix}^\intercal
$.

\subsubsection{$G_{11} = ({\bf a} \vert {\bf a}) > 0$ or $G_{22} = ({\bf b} \vert {\bf b}) > 0$, plane of type $(1,0)$}

In this subcase one of $({\bf a} \vert {\bf a})$ and $({\bf b} \vert {\bf b})$,
say $({\bf a} \vert {\bf a})$, is assumed to be positive. Then the other one,
in this case $({\bf b} \vert {\bf b})$, needs to be non-negative, because only
then the product $({\bf a} \vert {\bf a})({\bf b} \vert {\bf b})$ is
non-negative and therefore may ``compensate'' the subtraction of the
non-negative term $( {\bf b} \vert {\bf a})^2$ of the Gram
determinant~(\ref{2020-dex-e-gramian-determinant}). From \cite[Thm.~4,
p.~307]{Gantmacher1}, $\mathcal{M}$ is a plane of type $(1,0)$.

Decomposability~(\ref{2020-dex-e-di4dblfnccf}) requires that, for some $\xi$,
$({\xi}{\bf a}+{\bf b} \vert {\xi}{\bf a}+{\bf b}) = G_{11}\xi^2+2 G_{12}\xi +
G_{22}=0$, and thus, $\xi = \left( - 2 G_{12} \pm \sqrt{ 4 G_{12}^2 - 4 G_{11}
G_{22}}\right)/\left(2 G_{11}\right) = \Big( - G_{12} \pm \underbrace{\sqrt{ -
G} }_{=0}\Big)/G_{11} = - G_{11}^{-1}G_{12}$. Note that, in order for the
denominator $G_{11}=({\bf a} \vert {\bf a})$ not to vanish, $\xi$ must be
multiplied with the indecomposable vector ${\bf a}$. Therefore there exists (up
to scale factors) only a unique decomposable vector in $\mathcal{M}$, namely
${\bf c} = - G_{11}^{-1}G_{12}{\bf a}+{\bf b}$. All vectors in $\mathcal{M}$
that are not in the span of ${\bf c}$ -- indeed, a continuity of vectors -- are
indecomposable.

A typical example is the two-dimensional subspace spanned by
${\bf a}=
\begin{pmatrix}
0,1,-1,0
\end{pmatrix}^\intercal
$
and
$
{\bf b}=
\begin{pmatrix}
1,0,0,0
\end{pmatrix}^\intercal
$. Any element of the span of ${\bf a}$ and ${\bf b}$ can be written as
$
\begin{pmatrix}
x_1,x_2,-x_2,0
\end{pmatrix}^\intercal
$. The associated Gramian is of the form $ G_{{\bf a}{\bf b}}= \textrm{diag}
(2,0) $.

\subsubsection{$G_{11} = ({\bf a} \vert {\bf a}) < 0$ or $G_{22} = ({\bf b} \vert {\bf b}) < 0$, plane of type $(-1,0)$}

In this subcase one of $({\bf a} \vert {\bf a})$ and $({\bf b} \vert {\bf b})$,
say $({\bf a} \vert {\bf a})$, is assumed to be negative. Then the other one,
in this case $({\bf b} \vert {\bf b})$, needs to be non-positive, because only
then the product $({\bf a} \vert {\bf a})({\bf b} \vert {\bf b})$ is
non-positive and therefore may ``compensate'' the subtraction of the
non-negative term $( {\bf a} \vert {\bf b})^2$ of the Gram
determinant~(\ref{2020-dex-e-gramian-determinant}).

Again, there exists (up to scale factors) only a unique decomposable vector in
$\mathcal{M}$, namely ${\bf c} =  - G_{11}^{-1}G_{12}{\bf a}+{\bf b}$.

A typical example is the two-dimensional subspace spanned by
${\bf a}=
\begin{pmatrix}
0,1,1,0
\end{pmatrix}^\intercal
$
and
$
{\bf b}=
\begin{pmatrix}
1,0,0,0
\end{pmatrix}^\intercal
$. Any element of the span of ${\bf a}$ and ${\bf b}$ can be written as $
\begin{pmatrix}
x_1,x_2,x_2,0
\end{pmatrix}^\intercal
$.
The associated Gramian is of the form $ G_{{\bf a}{\bf b}}= \textrm{diag}
(-2,0) $.

\subsection{Gram determinant $G < 0$, plane of type $(1,-1)$}

In this case $({\bf a} \vert {\bf a})$ as well as $({\bf b} \vert {\bf b})$ can be
anything (positive, negative, zero). By the characterizations in~\cite[Thm.~3, 4, 5, 6, pp.~306--308]{Gantmacher1},
the plane $\mathcal{M}$ has to be of the only remaining type, that is, of type $(1,-1)$.

There exists (up to scale factors) only two unique distinct decomposable
vectors ${\bf c}_\pm$, in accordance with the construction given next. All
other vectors -- indeed, a continuity of vectors in the plane spanned by ${\bf
a}$ and ${\bf b}$ -- are indecomposable. This can again be shown by assuming
the case $({\bf a} \vert {\bf a})\neq 0$, and by noting that
decomposability~(\ref{2020-dex-e-di4dblfnccf}) requires that, for some $\xi$,
$({\xi}{\bf a}+{\bf b} \vert {\xi}{\bf a}+{\bf b}) = G_{11}\xi^2+2 G_{12}\xi +
G_{22}=0$, such that $\xi_\pm = \left( - 2 G_{12} \pm \sqrt{ 4 G_{12}^2 - 4
G_{11} G_{22}}\right)/\left(2 G_{11}\right) = \Big( - G_{12} \pm
\underbrace{\sqrt{ - G} }_{\neq 0}\Big)/G_{11} $. Note that these two solutions
${\bf c}_\pm = \left[\left( - G_{12} \pm \sqrt{ - G} \right)/G_{11}\right]{\bf
a}+{\bf b}$ need not be mutually orthogonal. In the second case one supposes
that, instead of $({\bf a} \vert {\bf a})\neq 0$, now $({\bf b}\vert {\bf
b})\neq 0$, and carries through an analogous calculation. In the third case
$({\bf a} \vert {\bf a})=({\bf b} \vert {\bf b})= 0$ both vectors ${\bf a}$ as
well as ${\bf b}$ are already decomposable. Note  that, in order for the
denominator not to vanish, $\xi$ must be multiplied with the respective
indecomposable vector.

A typical example is the two-dimensional subspace spanned by
${\bf a}=
\begin{pmatrix}
1,0,0,0
\end{pmatrix}^\intercal
$
and
$
{\bf b}=
\begin{pmatrix}
0,0,0,1
\end{pmatrix}^\intercal
$. Any element of the span of ${\bf a}$ and ${\bf b}$ can be written as $
\begin{pmatrix}
x_1,0,0,x_2
\end{pmatrix}^\intercal
$.
The associated Gramian is of the form
$
G_{{\bf a}{\bf b}}=
\begin{pmatrix}
0  & 1\\
1  & 0
\end{pmatrix}
$.

The main results of those considerations, as it concerns the question of
(in)decomposability, is that, with the exception of type~$(0,0)$ planes which
contain only decomposable vectors, all other five plane types contain (a
continuity of) orthogonal bases spanning them whose basis vectors are both
indecomposable: planes of type~$(1,-1)$ contain (up to scale factors) a single
orthogonal basis whose elements are decomposable; and planes of types~$(1,0)$
and~$(-1,0)$ contain (up to scale factors and permutations) a single orthogonal
basis with one decomposable and one indecomposable element. Planes of plane of
types~$(1,1)$ and~$(-1,-1)$ contain no decomposable non-zero vectors. This
completes the characterisation of the real case.

For complex four-dimensional Hilbert space $\mathcal{H}=\mathbb{C}^4$,
according to the cases~(ii.1) to~(ii.3) mentioned earlier, the rank of $G_{{\bf
a}{\bf b}}$ determines the type of $\mathcal{M}$ \cite[Theorem~11.24,
p.~287]{roman-advancedLA}:
\begin{itemize}

\item[(i)] If $\textrm{rank}(G_{{\bf a}{\bf b}}) = 0$ then $\mathcal{M}$ is
    of type $(0,0)$. Earlier remarks concerning properties of a plane of
    real type $(0,0)$ pertain.

\item[(ii)] If $\textrm{rank}(G_{{\bf a}{\bf b}}) = 1$ then $\mathcal{M}$
    is of type $(1,0)$. This situation parallels that of a plane of real
    type $(1,0)$. In particular, the calculation from there, yielding a
    unique decomposable vector ${\bf c}\in\mathcal{M}$, carries over
    provided that $G_{11}=({\bf a} \vert {\bf a})\neq 0$. Moreover, there
    is continuity of indecomposable vectors in $\mathcal{M}$ which are not
    in the span of ${\bf c}$.

\item[(iii)] If $\textrm{rank}(G_{{\bf a}{\bf b}}) = 2$ then $\mathcal{M}$
    is of type $(1,1)$. There is a neat analogy to the case of a plane of
    real type $(1,-1)$. Note that any quadratic equation over the complex
    numbers with non-vanishing discriminant has precisely two distinct
    solutions. Therefore, the calculation of the decomposable vectors ${\bf
    c_\pm}\in\mathcal{M}$ carries over, provided that $G_{11}=({\bf a}
    \vert {\bf a})\neq 0$.

\end{itemize}

All three plane types actually occur. This follows immediately from a
reinterpretation of our various examples in the real case.

The various types of planes admit a geometric interpretation in terms of the
\emph{projective space} $\mathbb{P}(\mathcal{H})$. We recall that the
\emph{points} of $\mathbb{P}(\mathcal{H})$ are the one-dimensional subspaces of
$\mathcal{H}$. A set of points is called a \emph{projective line}
(\emph{projective plane}) of $\mathbb{P}(\mathcal{H})$ if it comprises all
one-dimensional subspaces of $\mathcal{H}$ that are contained in some fixed
two-dimensional (three-dimensional) subspace of $\mathcal{H}$; see, for
example, \cite[p.~122]{odehnal+s+g}. All points of $\mathbb{P}(\mathcal{H})$
that are spanned by decomposable vectors constitute a \emph{ruled quadric}
$\Phi$, say, with equation $z_1z_4-z_2z_3 = 0$
\cite[pp.~143--144]{odehnal+s+g}. The \emph{type} of a projective line is
understood to be the type of the associated subspace of $\mathcal{H}$.

In the real case the points off the quadric $\Phi$ fall into two classes,
namely the sets of points $\mathrm{span}\{\bf{z}\}$, $\bf{z}\in \mathcal{H}$,
with $({\bf z} \vert {\bf z})>0$ and $({\bf z} \vert {\bf z})<0$, respectively.
We call these two classes the \emph{positive} and the \emph{negative side} of
$\Phi$, respectively. (From a geometric point of view, the attributes
``positive'' and ``negative'' are immaterial. Indeed, multiplying the equation
of $\Phi$ by some negative real number will change the labelling of the two
sides but not the quadric $\Phi$.) A projective line is of type $(0,0)$
precisely when it is contained in $\Phi$. A projective line is of type $(1,0)$
[of type $(-1,0)$] if and only if it meets $\Phi$ at a unique point whereas all
its other points are on the positive [negative] side of $\Phi$. The projective
lines of type $(1,1)$ [of type $(-1,-1)$] are those which are contained in the
positive [negative] side of $\Phi$. Finally, a projective line is of type
$(1,-1)$ precisely when it meets $\Phi$ at exactly two distinct points. (Any
such line contains points from either side.)

In the complex case, a projective line is of type $(0,0)$, $(1,0)$ or $(1,1)$
precisely when it is contained in $\Phi$, it meets $\Phi$ at a unique point or
it meets $\Phi$ at exactly two distinct points.

In order to visualize this situation in the real case, we consider the
\emph{affine space} on $\mathbb{R}^3$; its \emph{points} are the vectors of
$\mathbb{R}^3$, an \emph{affine line} (\emph{affine plane}) is a translate of a
one-dimensional (two-dimensional) subspace of $\mathbb{R}^3$. There is a
one-one correspondence between the set of points of $\mathbb{P}(\mathcal{H})$
that are not contained in the projective plane $z_4=0$ and the set of points of
the affine space on $\mathbb{R}^3$ as follows:
\begin{equation*}
    \mathrm{span}\{(z_1,z_2,z_3,z_4)^\intercal\}
    \mapsto (w_1,w_2,w_3)^\intercal=\left(\frac{z_1}{z_4},\frac{z_2}{z_4},\frac{z_3}{z_4}\right)^\intercal .
\end{equation*}
Under this correspondence a projective line (plane) corresponds to an affine
line (plane) unless it is contained in the projective plane $z_4=0$
\cite[p.~124]{odehnal+s+g}.
\par
\begin{figure}[htb]\unitlength=0.5\textwidth 
\centering
\begin{picture}(1,0.75)
\put(0,0){\includegraphics[width = 0.5\textwidth]{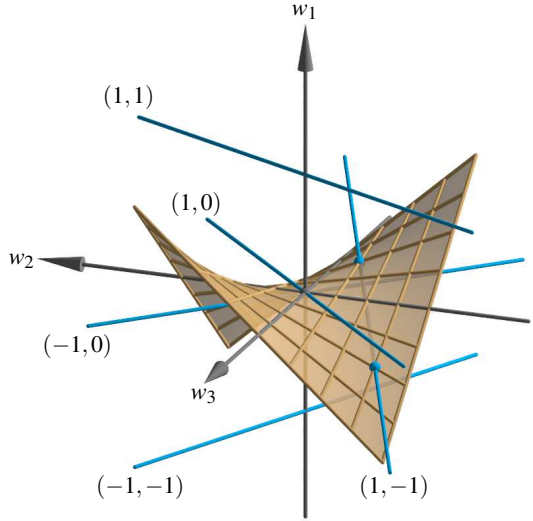}} 
    \put(0.06,0.38){$w_2$}
    \put(0.33,0.18){$w_3$}
    \put(0.48,0.75){$w_1$}
    \put(0.2,0.62){$(1,1)$}
    \put(0.3,0.46){$(1,0)$}
    \put(0.11,0.25){$(-1,0)$}
    \put(0.19,0.04){$(-1,-1)$}
    \put(0.58,0.04){$(1,-1)$}
\end{picture}
\caption{\label{2020-dec-fig-schema}
Schematic drawing of various plane types.
}
\end{figure}

The points (off the plane $z_4=0$) of the ruled quadric $\Phi$ correspond to
the points of a \emph{hyperbolic paraboloid} with equation $w_1=w_2 w_3$, which
is depicted in Fig.~\ref{2020-dec-fig-schema}. The figure also shows several
affine lines together with the type of their associated projective lines. All
affine lines on the paraboloid, among which are the $w_2$-axis and the
$w_3$-axis are of type $(0,0)$. The points ``above'' (''below'') the paraboloid
illustrate the positive (negative) side. The $w_1$-axis thereby is understood
to be ``tending upwards''. Take notice that this picture lacks all points of
$\mathbb{P}(\mathcal{H})$ in the projective plane $z_4=0$. Therefore, in some
cases, it provides an incomplete illustration. For example, the $w_1$-axis has
just one point in common with the paraboloid, namely $(0,0,0)^\intercal$ even
though it corresponds to a projective line of type $(1,-1)$, which is spanned
by the decomposable vectors $(1,0,0,0)^\intercal$ and $(0,0,0,1)^\intercal$.

The paraboloid from Fig.~\ref{2020-dec-fig-schema} is one way to visualize the ruled quadric $\Phi$
comprising all points that are spanned by decomposable vectors;
this can also be found in Refs.~\cite[Fig.~6, p.~4687]{Bengtsson2002} and \cite[Fig.~16.1, p.~438]{Bengtsson2017}.
For an alternative point of view, from which $\Phi$ appears as a hyperboloid of one sheet,
see  Refs.~\cite[Fig.~2.17, p.~35]{odehnal+s+g} and \cite[Fig.~4.3, p.~113]{Bengtsson2017}.

\section{Identification and characterization of (orthogonal) planes}

We are now in a position to solve the problem mentioned earlier: suppose we are
given two orthogonal unit vectors ${\bf e}_1$ and ${\bf e}_2$ spanning a plane
(aka the two-dimensional subspace) $\mathcal{M}$ of a four-dimensional Hilbert
space $\mathcal{H}$. One intermediate task -- a straightforward [e.g. via the
system of non-linear equations~(\ref{2020-e-unitary})] exercise -- is to find
an orthogonal basis $\left\{{\bf a},{\bf b}\right\}$ of the plane
$\mathcal{M}^\perp$ orthogonal to $\mathcal{M}$. Here we are not concerned with
the explicit realization of two remaining vectors ${\bf a}$ and ${\bf b}$. We
focus instead on the identification and analysis of the plane
$\mathcal{M}^\perp$, which allows us to decide whether or not ${\bf a}$ and
${\bf b}$ can be chosen indecomposable.

We shall solve this latter problem directly by substitution of the inner
product $\langle \cdot \vert \cdot \rangle$ by the bilinear form $( \cdot \vert
\cdot )$ from (\ref{2020-dex-e-di4dblf}), as exposed in
Eq.~(\ref{2020-dec-e-transcript}).

We start by introducing the plane which is defined as the image of
$\mathcal{M}$ under the (anti)linear
transformation~(\ref{2020-dec-def-x-tilde}):
\begin{equation}\label{hans-Mtilde}
\widetilde{\mathcal{M}} = \left\{ \left(\textsf{\textbf{A}}^{-1}\right)\cdot
\overline{{\bf x}} \mid {\bf x} \in \mathcal{M} \right\}
= \left\{ \widetilde{{\bf x}} \mid {\bf x} \in \mathcal{M} \right\}.
\end{equation}
The first essential point is as follows. \emph{The planes $\mathcal{M}$ and
$\widetilde{\mathcal{M}}$ are of the same type regarding $(\cdot\vert\cdot)$.}
In order to prove the assertion, we note that, by a straightforward
calculation,
\begin{equation}\label{2020-dec-umbenennen}
    (\widetilde{\bf x} \vert \widetilde{\bf y}) =
    \overline{({\bf x} \vert {\bf y})} \mbox{~~for all~~}
    {\bf x},{\bf y} \in \mathcal{H} .
\end{equation}
Now let $\{{\bf c}_1,{\bf c}_2\}$ be a basis of $\mathcal{M}$ such that the
Gramian matrix $G_{{\bf c}_1 {\bf c}_2}$ has the distinguished form as
described in \eqref{hans-real} for the real case or as in \eqref{hans-complex}
for the complex case. Then $\{\widetilde{\bf c}_1, \widetilde{\bf c}_2\}$ is a
basis of $\widetilde{\mathcal{M}}$ and \eqref{2020-dec-umbenennen} gives
\begin{equation}\label{2020-dec-gram-gram}
    G_{\widetilde{\bf c}_1 \widetilde{\bf c}_2} =
    \overline {G_{{\bf c}_1{\bf c}_2}} =
    {G_{{\bf c}_1{\bf c}_2}} ,
\end{equation}
which establishes the result.

The second essential point is that the {\em type of $\mathcal{M}^\perp$
regarding $(\cdot\vert\cdot)$ is co-determined by the type of
$\widetilde{\mathcal{M}}$ regarding $(\cdot\vert\cdot)$.} Notice, however,
that, as earlier, the real and complex cases have to be treated separately:
whereas in the complex (Hilbert space) case $\mathcal{M}^\perp $ and
$\widetilde{\mathcal{M}}$ are of the same type, in the real (Hilbert space)
case

\begin{itemize}

\item[(i)]  $\mathcal{M}^\perp $ is of type   $(0,0)$ $\Leftrightarrow$ $\widetilde{\mathcal{M}}$ is of type   $(0,0)$,

\item[(ii)] $\mathcal{M}^\perp $ is of type   $(1,-1)$ $\Leftrightarrow$ $\widetilde{\mathcal{M}}$ is of type   $(1,-1)$,

\item[(iii)] $\mathcal{M}^\perp $ is of type $( {\pm}1, {\pm}1)$
    $\Leftrightarrow$ $\widetilde{\mathcal{M}}$ is of type
    $( {\mp}1, {\mp}1)$,

\item[(iv)] $\mathcal{M}^\perp $ is of type $({\pm}1,0)$
    $\Leftrightarrow$ $\widetilde{\mathcal{M}}$ is of type
    $( {\mp}1,0)$.

\end{itemize}

Our proof is based on the following alternative description of
$\widetilde{\mathcal {M}}$, which makes use of Eq.~\eqref{hans-Mtilde}, the
identity $\mathcal{M} = (\mathcal{M}^\perp)^\perp$,
Eq.~(\ref{2020-dex-e-di4dblf}) and the bijectivity of the
transformation~(\ref{2020-dec-def-x-tilde}):
\begin{equation}\label{hans-ortho1}
\begin{aligned}
    \widetilde{\mathcal {M}} &= \{\widetilde{\bf x}\in \mathcal{H} \mid {\bf x}\in \mathcal{M}\}\\
                             &= \{\widetilde{\bf x}\in \mathcal{H} \mid {\bf x}\in(\mathcal{M}^\perp)^\perp\}\\
                             &= \{\widetilde{\bf x}\in \mathcal{H} \mid \langle{\bf x}\vert{\bf y}\rangle = 0 \mbox{~for all~} {\bf y}\in\mathcal{M}^\perp\}\\
                             &= \{\widetilde{\bf x}\in \mathcal{H} \mid (\widetilde{\bf x}\vert{\bf y}) = 0 \mbox{~for all~} {\bf y}\in\mathcal{M}^\perp\}\\
                             &= \{          {\bf z}\in \mathcal{H} \mid (          {\bf z}\vert{\bf y}) = 0 \mbox{~for all~} {\bf y}\in\mathcal{M}^\perp\}.
\end{aligned}
\end{equation}
Likewise, we also have
\begin{equation}\label{hans-ortho2}
\begin{aligned}
    \mathcal{M}^\perp & = \{{\bf y}\in \mathcal{H} \mid \langle {\bf x}\vert{\bf y}\rangle = 0 \mbox{~for all~} {\bf x}\in\mathcal{M}\}\\
                      & = \{{\bf y}\in \mathcal{H} \mid (\widetilde{\bf x}\vert{\bf y}) = 0    \mbox{~for all~} {\bf x}\in\mathcal{M}\}\\
                      & = \{{\bf y}\in \mathcal{H} \mid ({\bf z}\vert{\bf y}) = 0              \mbox{~for all~} {\bf z}\in\widetilde{\mathcal{M}}\} .
\end{aligned}
\end{equation}
Eqs.~\eqref{hans-ortho1} and \eqref{hans-ortho2} imply that
\begin{equation}\label{hans-radical}
    \mathrm{rad}(\widetilde{\mathcal{M}}) =
    \widetilde{\mathcal{M}} \cap \mathcal{M}^\perp =
    \mathrm{rad}(\mathcal{M}^\perp) .
\end{equation}
There exist bases $\{{\bf d}_1,{\bf d}_2\}$ of $\widetilde{\mathcal{M}}$ and
$\{{\bf d}_3,{\bf d}_4\}$ of $\mathcal{M}^\perp$ such that their Gramian
matrices have the distinguished form as described in \eqref{hans-real} for the
real case or as in \eqref{hans-complex} for the complex case. Let $m$ denote
the dimension of the subspace appearing in Eq.~\eqref{hans-radical}. Then
Eq.~\eqref{hans-radical=span}, applied to $\widetilde{\mathcal{M}}$ and its
basis $\{{\bf d}_1,{\bf d}_2\}$, together with one of Eqs.~\eqref{hans-real}
and \eqref{hans-complex} shows that the leading $2-m$ diagonal entries of the
Gramian matrix $G_{{\bf d}_1{\bf d}_2}$ are non-zero, whereas the remaining $m$
diagonal entries are zero. The same result holds, mutatis mutandis, for the
Gramian matrix $G_{{\bf d}_3{\bf d}_4}$. There are three cases.

In the first case, $\mathcal {H}$ is a complex Hilbert space or $m=2$. Then, by
the above, $G_{{\bf d}_1{\bf d}_2} = G_{{\bf d}_3{\bf d}_4}$ so that
$\widetilde{\mathcal{M}}$ and $\mathcal{M}^\perp$ are of the same type. In
particular, for $m=2$ both planes are of type $(0,0)$. This establishes the
result for a complex space as well as (i) for a real space.

In the second case, $\mathcal {H}$ is a real Hilbert space and $m=0$. The
planes $\widetilde{\mathcal{M}}$ and $\mathcal{M}^\perp$ are of types
$(\epsilon_1,\epsilon_2)$ and $(\epsilon_3,\epsilon_4)$, respectively, where
$\epsilon_1,\epsilon_2,\epsilon_3,\epsilon_4\in\{1,-1\}$. Since
$m=\dim(\widetilde{\mathcal{M}} \cap \mathcal{M}^\perp)=0$, the four
vectors ${\bf d}_1,{\bf d}_2,{\bf d}_3,{\bf d}_4 $ constitute a basis
of $\widetilde{\mathcal{M}} \oplus \mathcal{M}^\perp = \mathcal{H}$ and
$G_{{\bf d}_1{\bf d}_2{\bf d}_3{\bf
d}_4}=\mathrm{diag}\left(\epsilon_1,\epsilon_3,\epsilon_3,\epsilon_4\right)$.
By Sylvester's law of inertia \cite[Thm.~1, p.~297]{Gantmacher1}, this
matrix coincides -- up to a permutation of its diagonal entries -- with
the matrix $\textsf{\textbf{A}}'=\textrm{diag}(1,1,-1,-1)$ from
Section~II. This establishes (ii) and (iii).

In the third case, $\mathcal {H}$ is a real Hilbert space and $m=1$. The planes
$\widetilde{\mathcal{M}}$ and $\mathcal{M}^\perp$ are of types $(\epsilon_1,0)$
and $(\epsilon_3,0)$, respectively, where $\epsilon_1,\epsilon_3\in\{1,-1\}$.
In order to verify (iv), it remains to show that $\epsilon_1$ and $\epsilon_3$
have different signs. Assume to the contrary that, for example, $\epsilon_1$
and $\epsilon_3$ are both positive. From Eq.~\eqref{hans-radical=span}, applied
to $\widetilde{\mathcal{M}}$ and its basis $\{{\bf d}_1,{\bf d}_2\}$, we obtain
${\bf d}_1\notin \mathrm{rad}(\widetilde{\mathcal{M}})$ and ${\bf d}_2 \in
\mathrm{rad}(\widetilde{\mathcal{M}})$. The same kind of reasoning for
$\mathcal{M}^\perp$ and $\{{\bf d}_3,{\bf d}_4\}$ yields ${\bf d}_3\notin
\mathrm{rad}(\mathcal{M}^\perp)$ and ${\bf d}_4 \in
\mathrm{rad}(\mathcal{M}^\perp)$. Thus, using Eq.~\eqref{hans-radical}, ${\bf
d}_1,{\bf d}_3 \notin \widetilde{\mathcal{M}} \cap \mathcal{M}^\perp$ whereas
${\bf d}_2,{\bf d}_4 \in \widetilde{\mathcal{M}} \cap \mathcal{M}^\perp$. The
three vectors ${\bf d}_1,{\bf d}_2,{\bf d}_3$ therefore constitute a basis of
$\widetilde{\mathcal{M}} + \mathcal{M}^\perp $ and its Gramian matrix has the
form $G_{{\bf d}_1{\bf d}_2{\bf d}_3} =
\mathrm{diag}\left(\epsilon_1,0,\epsilon_3\right)=\mathrm{diag}\left(1,0,1\right)$.
Therefore $({\bf x} \vert {\bf x})\geq 0$ for all ${\bf x} \in
\widetilde{\mathcal{M}} + \mathcal{M}^\perp$. On the other hand, there exists a
plane $\mathcal{N}$ of type $(-1,-1)$, whence $({\bf x} \vert {\bf x}) < 0$ for
all non-zero vectors ${\bf x} \in \mathcal{N}$. Due to $\dim \mathcal{H}=4$,
the plane $\mathcal{N}$ has a non-zero intersection with the three-dimensional
subspace $\widetilde{\mathcal{M}}+\mathcal{M}^\perp $, that is, there exists a
vector ${\bf n}\in \widetilde{\mathcal{M}} + \mathcal{M}^\perp $ with $({\bf
n},{\bf n})<0$, a contradiction.

Summing up, the plane type of $\mathcal{M}^\perp $ can be directly obtained by
analyzing the Gramian matrix $G_{{\bf e}_1 {\bf e}_2}$ of the two given
``input'' vectors ${\bf e}_1$ and $ {\bf e}_2$, and, in the complex case,
determining its rank.

\section{Orthogonality of decomposable vectors}\label{se:V}

Throughout this section, $\mathcal{M}$ denotes a plane of type $(1,-1)$ (real
case) or of type $(1,1)$ (complex case). Then there exist vectors ${\bf s},{\bf
t},{\bf u},{\bf v}$ in $\mathcal{H}_2$ such that
\begin{equation}\label{hans-M-basis}
    \bigl\{{\bf s}\otimes{\bf t}, {\bf u}\otimes{\bf v}\bigr\} ,
\end{equation}
is a basis of $\mathcal{M}$. Since $\mathcal{M}$ is not of type $(0,0)$, we
must have, by virtue of Eq.~\eqref{hans-bilinear-det},
\begin{equation}\label{hans-nicht0}
    ({\bf s}\otimes{\bf t}\vert{\bf u}\otimes{\bf v})
    =\det({\bf s},{\bf u}) \det({\bf t},{\bf v}) \neq 0 .
\end{equation}
This in turn shows that $\{{\bf s}, {\bf u}\} $ and $\{{\bf t}, {\bf v}\}$ are
bases of $\mathcal{H}_2$. Now \eqref{hans-nicht0} implies $\det({\bf
s}^\times,{\bf u}^\times)=\overline{\det({\bf s},{\bf u})}\neq 0$ and
$\det({\bf t}^\times,{\bf v}^\times)=\overline{\det({\bf t},{\bf v})}\neq 0$.
Therefore each of the sets $\{{\bf s}^\times,{\bf u}^\times\}$ and $\{{\bf
t}^\times,{\bf v}^\times\}$ is a basis of $\mathcal{H}_2$. Consequently, we
obtain
\begin{equation}\label{hans-Tensorbasis}
\bigl\{         {\bf s}^\times\otimes{\bf t}^\times,\;
    \underbrace{{\bf s}^\times\otimes{\bf v}^\times}_{=:\,{\bf b}},\;
    \underbrace{{\bf u}^\times\otimes{\bf t}^\times}_{=:\,{\bf a}},\;
                {\bf u}^\times\otimes{\bf v}^\times
\bigr\}
\end{equation}
as a basis of $\mathcal{H}$. Furthermore, it is immediate from
Eqs.~\eqref{hans-inner-rel} and \eqref{hans-orthogonal} that each of the
linearly independent vectors ${\bf a}$ and ${\bf b}$ is orthogonal to the
vectors ${\bf s}\otimes{\bf t}$ and ${\bf u}\otimes{\bf v}$, that is, $\{{\bf
a},{\bf b}\}$ is a basis of the plane $\mathcal{M}^\perp$.

If the basis vectors of $\mathcal{M}$ appearing in \eqref{hans-M-basis} are
orthogonal, may we then suspect that there exists a ``completed'' orthogonal
basis of the four-dimensional real or complex Hilbert space ${\cal H}$ which
(includes these two vectors and) consists solely of decomposable vectors?
Stated pointedly, does the orthogonality of decomposable vectors spanning the
given plane $\mathcal{M}$ imply that the corresponding two decomposable vectors
in the orthogonal subspace $\mathcal{M}^\perp$ [which is again of the same
type] are also orthogonal, and {\it vice versa?} In what follows we shall prove
that this is indeed the case; that is, the orthogonality of the two
decomposable vectors from \eqref{hans-M-basis} is ``inherited'' by the two
decomposable vectors ${\bf a}$ and ${\bf b}$ defined in
\eqref{hans-Tensorbasis}. Using Eqs.~\eqref{hans-inner-rel} and
\eqref{hans-inner-x} we obtain:
\begin{equation}\label{hans-inner-M}
    \langle{\bf s}\otimes{\bf t} \vert {\bf u}\otimes{\bf v}\rangle =
    \langle{\bf s}\vert{\bf u}\rangle \langle{\bf t}\vert{\bf v}\rangle
\end{equation}
and
\begin{equation}\label{hans-inner-M-perp}
    \langle{\bf u}^\times\otimes{\bf t}^\times \vert {\bf s}^\times\otimes{\bf v}^\times\rangle =
    \langle{\bf u}^\times\vert{\bf s}^\times\rangle \langle{\bf t}^\times\vert{\bf v}^\times\rangle =
    \langle{\bf s}\vert{\bf u}\rangle \langle{\bf v}\vert{\bf t}\rangle .
\end{equation}
Therefore ${\bf s}\otimes{\bf t}$ and ${\bf u}\otimes{\bf v}$ are orthogonal if
and only if at least one of the inner products $\langle{\bf s}\vert{\bf
u}\rangle$ and $\langle{\bf t}\vert{\bf v}\rangle$ vanishes. This in turn is
equivalent to ${\bf u}^\times\otimes{\bf t}^\times$ and ${\bf
s}^\times\otimes{\bf v}^\times$ being orthogonal.

Our considerations from above do not involve the auxiliary plane
$\widetilde{\mathcal{M}}$ that we used before. We add, for the sake of
completeness, that a basis of $\widetilde{\mathcal{M}}$ is given by
\begin{equation}\label{hans-Mtilde-basis}
    \bigl\{ {\bf s}^\times\otimes{\bf t}^\times ,\;
            {\bf u}^\times\otimes{\bf v}^\times
    \bigr\} .
\end{equation}
This follows from Eq.~\eqref{hans-tilde-times} applied to the basis vectors of
$\mathcal{M}$ from \eqref{hans-M-basis}. Note that Eq.~(\ref{hans-inner-M}) and
the analogue of (\ref{hans-inner-M-perp}) (obtained by interchanging ${\bf
s}^\times$ and ${\bf u}^\times$) establishes that the orthogonality of the
decomposable basis vectors of $\widetilde{\mathcal{M}}$ from
\eqref{hans-Mtilde-basis} holds precisely when the decomposable basis vectors
of $\mathcal{M}$ from \eqref{hans-M-basis} are orthogonal.

All things considered, we see that the four decomposable basis vectors from
\eqref{hans-Tensorbasis} give rise to six planes. Two of them are
$\mathcal{M}^\perp$ and $\widetilde{\mathcal{M}}$. The remaining four planes
are of type $(0,0)$: Take, for example, the plane spanned by ${\bf
s}^\times\otimes{\bf t}^\times$ and ${\bf s}^\times\otimes{\bf v}^\times$. An
arbitrary linear combination of these two vectors reads $\xi_1({\bf
s}^\times\otimes{\bf t}^\times)+\xi_2({\bf s}^\times\otimes{\bf v}^\times) =
{\bf s}^\times\otimes(\xi_1{\bf t}^\times +\xi_2{\bf v}^\times)$ and therefore
is decomposable.

\begin{figure}[htb]\unitlength=0.5\textwidth 
\centering
\begin{picture}(1,0.75)
\put(0,0){\includegraphics[width = 0.5\textwidth]{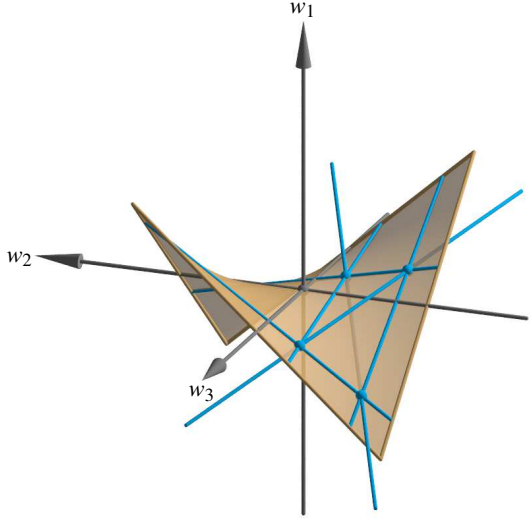}} 
    \put(0.06,0.38){$w_2$}
    \put(0.33,0.18){$w_3$}
    \put(0.48,0.75){$w_1$}
\end{picture}
\caption{\label{2020-dec-fig-tetra}
Tetrahedron arising from the basis \eqref{hans-Tensorbasis}.
}
\end{figure}

In a geometric language, the four vectors from \eqref{hans-Tensorbasis}
generate the vertices of a \emph{tetrahedron} with three specific properties in
the projective space $\mathbb{P}(\mathcal{H})$. First, the vertices of the
tetrahedron are on the ruled quadric $\Phi$, whose points are given by all
non-zero decomposable vectors. Second, two edges of the tetrahedron meet $\Phi$
at exactly two distinct points. Third, the remaining four edges lie completely
on the quadric $\Phi$. One tetrahedron of this kind is depicted in
Fig.~\ref{2020-dec-fig-tetra}, where we adopted the same affine viewpoint as in
Fig.~\ref{2020-dec-fig-schema}.

It is a straightforward task to obtain \emph{all} planes of type $(1,-1)$ (real
case) and of type $(1,1)$ (complex case) by a reverse approach. Given any two
bases $\{{\bf s}', {\bf u}'\} $ and $\{{\bf t}', {\bf v}'\}$ of $\mathcal{H}_2$
the analogue of \eqref{hans-nicht0} holds. This shows that the plane spanned by
${\bf s}'\otimes {\bf t}'$ and ${\bf u}'\otimes {\bf v}'$ has the required
type. Furthermore, by an appropriate choice of the initial bases, one can
assure that ${\bf s}'\otimes {\bf t}'$ and ${\bf u}'\otimes {\bf v}'$ are
(non-)orthogonal.

\section{Consequences for completion of contexts}

One of the main results of this categorization exercise is that, as long as the
two given vectors ${\bf e}_1$ and ${\bf e}_2$ do not span a plane of
type~$(0,0)$ -- that is, as long as their Gramian does not vanish such that
$G_{{\bf e}_1 {\bf e}_2} = \textrm{diag} (0,0) $ -- the vectors completing the
context (four-dimensional orthogonal basis) can always be chosen to be
indecomposable, and therefore correspond to entangled states. In the case the
Gramian $G_{{\bf e}_1 {\bf e}_2}$ vanishes the entire context consists of
decomposable vectors associated with non-entangled states. Moreover, if there
exist two orthogonal decomposable vectors spanning a plane it is always
possible to ``complete'' the respective orthogonal basis by adding two
orthogonal decomposable vectors spanning the orthogonal subspace.

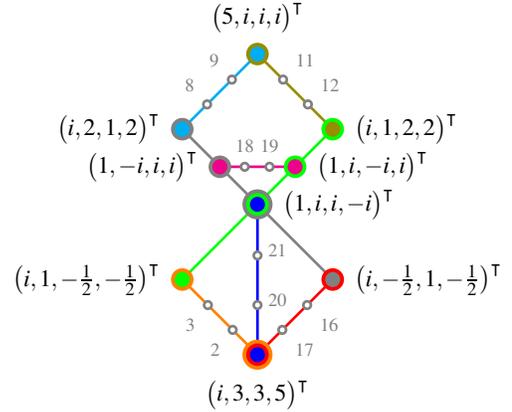
\begin{figure}[htb]
\begin{center}
\begin{tikzpicture}  [scale=0.5]

\tikzstyle{every path}=[line width=1pt]

\newdimen\ms
\ms=0.1cm
\tikzstyle{s1}=[color=red,rectangle,inner sep=3.5]
\tikzstyle{c3}=[circle,inner sep={\ms/8},minimum size=4*\ms]
\tikzstyle{c2}=[circle,inner sep={\ms/8},minimum size=3*\ms]
\tikzstyle{c1}=[circle,inner sep={\ms/8},minimum size=2*\ms]
\tikzstyle{cs1}=[circle,inner sep={\ms/8},minimum size=1*\ms]


\coordinate (psi) at (2,0);
\coordinate (vd) at (0,2);
\coordinate (dv) at (4,2);
\coordinate (uu) at (2,4);
\coordinate (vu) at (1,5);
\coordinate (uv) at (3,5);
\coordinate (cv) at (0,6);
\coordinate (vc) at (4,6);
\coordinate (dd) at (2,8);


\draw [color=orange] (psi) -- (vd)  coordinate[cs1,fill=white,draw=gray,pos=0.33,label=below left:{\scriptsize \color{gray}2}] (2)  coordinate[cs1,fill=white,draw=gray,pos=0.66,label=below left:{\scriptsize \color{gray}3}] (3);
\draw [color=blue] (psi) -- (uu)  coordinate[cs1,fill=white,draw=gray,pos=0.33,label={[right,xshift=0.2mm]:{\scriptsize \color{gray}20}}] (20)  coordinate[cs1,fill=white,draw=gray,pos=0.66,label={[right,xshift=0.2mm]:{\scriptsize \color{gray}21}}] (21);
\draw [color=red] (psi) -- (dv) coordinate[cs1,fill=white,draw=gray,pos=0.33,label=below right:{\scriptsize \color{gray}17}]  (17) coordinate[cs1,fill=white,draw=gray,pos=0.66,label=below right:{\scriptsize \color{gray}16}] (16);
\draw [color=green] (vd) -- (vc);
\draw [color=gray] (dv) -- (cv);
\draw [color=magenta] (vu) -- (uv) coordinate[cs1,fill=white,draw=gray,pos=0.33,label=above:{\scriptsize \color{gray}18}] (18)  coordinate[cs1,fill=white,draw=gray,pos=0.66,label=above:{\scriptsize \color{gray}19}] (19);
\draw [color=cyan] (cv) -- (dd) coordinate[cs1,fill=white,draw=gray,pos=0.33,label=above left:{\scriptsize \color{gray}8}]  (8) coordinate[cs1,fill=white,draw=gray,pos=0.66,label=above left:{\scriptsize \color{gray}9}] (9);
\draw [color=olive] (vc) -- (dd) coordinate[cs1,fill=white,draw=gray,pos=0.33,label=above right:{\scriptsize \color{gray}12}] (12)  coordinate[cs1,fill=white,draw=gray,pos=0.66,label=above right:{\scriptsize \color{gray}11}] (12);


\draw (psi)
coordinate[c3,fill=orange,label=below:{$\begin{pmatrix}i,3,3,5\end{pmatrix}^\intercal$}];
\draw (psi) coordinate[c2,fill=red]; \draw (psi) coordinate[c1,fill=blue];

\draw (vd)
coordinate[c2,fill=orange,label=left:{$\begin{pmatrix}i,1,-\frac{1}{2},-\frac{1}{2}\end{pmatrix}^\intercal$}];
\draw (vd) coordinate[c1,fill=green];

\draw (dv)
coordinate[c2,fill=red,label=right:{$\begin{pmatrix}i,-\frac{1}{2},1,-\frac{1}{2}\end{pmatrix}^\intercal$}];
\draw (dv) coordinate[c1,fill=gray];

\draw (uu)
coordinate[c3,fill=gray,label=right:{$\begin{pmatrix}1,i,i,-i\end{pmatrix}^\intercal$}];
\draw (uu) coordinate[c2,fill=green]; \draw (uu) coordinate[c1,fill=blue];

\draw (vu)
coordinate[c2,fill=gray,label=left:{$\begin{pmatrix}1,-i,i,i\end{pmatrix}^\intercal$}];
\draw (vu) coordinate[c1,fill=magenta];

\draw (uv)
coordinate[c2,fill=green,label=right:{$\begin{pmatrix}1,i,-i,i\end{pmatrix}^\intercal$}];
\draw (uv) coordinate[c1,fill=magenta];

\draw (cv)
coordinate[c2,fill=gray,label=left:{$\begin{pmatrix}i,2,1,2\end{pmatrix}^\intercal$}];
\draw (cv) coordinate[c1,fill=cyan];

\draw (vc)
coordinate[c2,fill=green,label=right:{$\begin{pmatrix}i,1,2,2\end{pmatrix}^\intercal$}];
\draw (vc) coordinate[c1,fill=olive];

\draw (dd)
coordinate[c2,fill=olive,label=above:{$\begin{pmatrix}5,i,i,i\end{pmatrix}^\intercal$}];
\draw (dd) coordinate[c1,fill=cyan];

\end{tikzpicture}
\end{center}
\caption{\label{2020-dec-hardy-fig1}
``Incomplete'' faithful orthogonal representation (aka coordinatization)
of the orthogonality hypergraph of
 the Hardy gadget, as quoted from Figure~1 and the last row of Table~I of Ref.~\cite{svozil-2020-hardy}.}
\end{figure}

For the sake of a concrete example consider the faithful orthogonal
representation (aka coordinatization) of a hypergraph of the Hardy type, as
quoted from the last row of Table~I of Ref.~\cite{svozil-2020-hardy}, as
depicted in Fig.~\ref{2020-dec-hardy-fig1}.

It comprises a pasting of two complete (aka orthogonal bases, maximal
operators~\cite[\S~84, Theorem~1]{halmos-vs}, Boolean subalgebras or
blocks~\cite{Greechie1968}, maximal cliques)
as well as six incomplete intertwining contexts
\begin{equation}
\begin{aligned}
&\left\{  \begin{pmatrix}i,1,-\frac{1}{2},-\frac{1}{2}\end{pmatrix}^\intercal,\begin{pmatrix}i,3,3,5\end{pmatrix}^\intercal ,2,3  \right\},  \\
&\left\{  \begin{pmatrix}5,i,i,i\end{pmatrix}^\intercal, \begin{pmatrix}i,2,1,2\end{pmatrix}^\intercal   ,8,9\right\}, \\
&\left\{  \begin{pmatrix}5,i,i,i\end{pmatrix}^\intercal, \begin{pmatrix}i,1,2,2\end{pmatrix}^\intercal ,11,12  \right\},\\
&\left\{  \begin{pmatrix}i,-\frac{1}{2},1,-\frac{1}{2}\end{pmatrix}^\intercal,\begin{pmatrix}i,3,3,5\end{pmatrix}^\intercal ,16,17 \right\},\\
&\left\{  \begin{pmatrix}1,-i,i,i\end{pmatrix}^\intercal, \begin{pmatrix}1,i,-i,i\end{pmatrix}^\intercal ,18,19  \right\},\\
&\left\{  \begin{pmatrix}1,i,i,-i\end{pmatrix}^\intercal,\begin{pmatrix}i,3,3,5\end{pmatrix}^\intercal ,20,21  \right\},
\end{aligned}
\label{2020-dex-hardy-contexts-in}
\end{equation}
arranged in  and 21 atoms or vectors, $2 \times 6=12$ thereof undefined, namely (partitions indicate same contexts)
$\{
\{2,3\},\{8,9\},\{11,12\},\{16,17\},\{18,19\},\{20,21\}
\}$.

By now it should be clear that all of these undefined vectors can be made to be indecomposable:
by a parity argument using their even numbers of imaginary units all of the defined vectors are indecomposable;
hence there is no way that these could  span a (transformed) type~$(0,0)$ plane.
But in what plane types exactly are those undefined vectors?
All we need to know is the type of the planes spanned by the transformed known vector pairs,
which reduces to the task of computing the rank of their Gramian matrices.

For the sake of an explicit computation, take the context defined by $\left\{
\begin{pmatrix}i,1,-\frac{1}{2},-\frac{1}{2}\end{pmatrix}^\intercal,\begin{pmatrix}i,3,3,5\end{pmatrix}^\intercal
,2,3 \right\}$, and identify ${\bf e}_1 =
\begin{pmatrix}i,1,-\frac{1}{2},-\frac{1}{2}\end{pmatrix}^\intercal$, ${\bf e}_2 =
\begin{pmatrix}i,3,3,5\end{pmatrix}^\intercal$, ${\bf a}= 2$, ${\bf b}= 3$,
respectively.
Then the associated Gramian matrix is $G_{{\bf e}_1 {\bf e}_2}=
\frac{1}{4}\begin{pmatrix}
 \XX 2-2 i & -3+9 i \\
 -3+9 i & -36+20 i
\end{pmatrix}
$.
The rank of this matrix is two; therefore the type of plane spanned by the
vectors ${\bf a}= 2$ and ${\bf b}= 3$ is $(1,1)$. Analogous computations show
that all planes spanned by the ``missing'' vectors are of type $(1,1)$.

Intuitively speaking there exist ``much less'' decomposable vectors than
indecomposable ones: from all vectors of four-dimensional space only those
satisfying condition~(\ref{2020-dex-e-di4dblfnccf}) qualify. Therefore, the
task of finding a faithful orthogonal representation \emph{with only
decomposable vectors} of a (hyper)graph turns out to be more difficult than,
say, by requiring indecomposability of the vectors. For some configurations and
(hyper)graphs it is impossible to find faithful orthogonal representations by
decomposable vectors; even if there exist ``plenty'' of such representations
containing also indecomposable vectors.

\begin{figure}[htb]
\begin{center}
\begin{tabular}{cc}
\begin{tikzpicture}  [scale=1]

        \tikzstyle{every path}=[line width=1pt]

        \newdimen\ms
        \ms=0.1cm \tikzstyle{s1}=[color=red,rectangle,inner sep=3.5]
        \tikzstyle{c3}=[circle,inner sep={\ms/8},minimum size=4*\ms]
        \tikzstyle{c2}=[circle,inner sep={\ms/8},minimum size=3*\ms]
        \tikzstyle{c1}=[circle,inner sep={\ms/8},minimum size=2*\ms]
        \tikzstyle{cs1}=[circle,inner sep={\ms/8},minimum size=1*\ms]


        \newdimen\alength
        \alength=2cm

        \coordinate (st) at (0,{\alength * sqrt(3)/2}); \coordinate (stx) at
        ({-\alength / 2},0); \coordinate (sxt) at ({\alength / 2},0);

        \draw [color=blue] (st) -- (stx)
        coordinate[c1,fill=blue,pos=0.33,label={[left,xshift=-0.2mm]:{\color{black}${\bf
        b}_2$}}] (sxu)
        coordinate[c1,fill=blue,pos=0.66,label={[left,xshift=-0.2mm]:{\color{black}${\bf
        b}_3$}}] (sxux); \draw [color=red] (st) -- (sxt)
        coordinate[c1,fill=red,pos=0.33,label={[right,xshift=0.2mm]:{\color{black}${\bf
        b}_5$}}] (vtx)
        coordinate[c1,fill=red,pos=0.66,label={[right,xshift=0.2mm]:{\color{black}${\bf
        b}_6$}}] (vxtx); \draw [color=green] (stx) -- (sxt)
        coordinate[c1,fill=green,pos=0.33,label={[below,xshift=-0.0mm,yshift=-1.5mm]:{\color{black}${\bf
        b}_8$}}] (sxtx)
        coordinate[c1,fill=green,pos=0.66,label={[below,xshift=-0.0mm,yshift=-1.5mm]:{\color{black}${\bf
        b}_9$}}] (xyz);


        \draw (st) coordinate[c2,fill=blue,label=above:{\color{black}${\bf
        b}_1$}]; \draw (st) coordinate[c1,fill=red];

        \draw (stx)
        coordinate[c2,fill=blue,label={[left,xshift=-0.4mm,yshift=-1.2mm]:\color{black}${\bf
        b}_4$}]; \draw (stx) coordinate[c1,fill=green];

        \draw (sxt)
        coordinate[c2,fill=red,label={[right,xshift=0.8mm,yshift=-1.2mm]:\color{black}${\bf
        b}_7$}]; \draw (sxt) coordinate[c1,fill=green];

        \end{tikzpicture}
        &
        \begin{tikzpicture}  [scale=1]

        \tikzstyle{every path}=[line width=1pt]

        \newdimen\ms
        \ms=0.1cm \tikzstyle{s1}=[color=red,rectangle,inner sep=3.5]
        \tikzstyle{c3}=[circle,inner sep={\ms/8},minimum size=4*\ms]
        \tikzstyle{c2}=[circle,inner sep={\ms/8},minimum size=3*\ms]
        \tikzstyle{c1}=[circle,inner sep={\ms/8},minimum size=2*\ms]
        \tikzstyle{cs1}=[circle,inner sep={\ms/8},minimum size=1*\ms]


        \newdimen\alength
        \alength=2cm

        \coordinate (st) at (0,{\alength * sqrt(3)/2}); \coordinate (stx) at
        ({-\alength / 2},0); \coordinate (sxt) at ({\alength / 2},0);


        \draw [color=blue] (st) -- (stx)
        coordinate[c1,fill=blue,pos=0.33,label={[left,xshift=-0.2mm]:{\color{black}${\bf
        s}^\times \otimes {\bf v}$}}] (sxu)
        coordinate[c1,fill=blue,pos=0.66,label={[left,xshift=-0.2mm]:{\color{black}${\bf
        s}^\times \otimes {\bf v}^\times$}}] (sxux); \draw [color=red] (st) --
        (sxt)
        coordinate[c1,fill=red,pos=0.33,label={[right,xshift=0.2mm]:{\color{black}${\bf
        u} \otimes {\bf t}^\times$}}] (vtx)
        coordinate[c1,fill=red,pos=0.66,label={[right,xshift=0.2mm]:{\color{black}${\bf
        u}^\times \otimes {\bf t}^\times$}}] (vxtx); \draw [color=green] (stx)
        -- (sxt)
        coordinate[c1,fill=green,pos=0.33,label={[below,xshift=-1.0mm,yshift=-1.2mm]:{\color{black}${\bf
        s}^\times \otimes {\bf t}^\times$}}] (sxtx)
        coordinate[c1,fill=green,pos=0.66,label={[below,xshift=1.4mm,yshift=-1.2mm]:{\color{black}${\bf
        s} \otimes {\bf t}\vphantom{^\times}$}}] (xyz); 


        \draw (st) coordinate[c2,fill=blue,label=above:{\color{black}${\bf
        s}\otimes {\bf t}$}]; \draw (st) coordinate[c1,fill=red];

        \draw (stx)
        coordinate[c2,fill=blue,label={[left,xshift=-0.4mm,yshift=-1.2mm]:\color{black}${\bf
        s}\otimes {\bf t}^\times$}]; \draw (stx) coordinate[c1,fill=green];

        \draw (sxt)
        coordinate[c2,fill=red,label={[right,xshift=0.8mm,yshift=-1.2mm]:\color{black}${\bf
        s}^\times\otimes {\bf t}$}]; \draw (sxt) coordinate[c1,fill=green];

        \end{tikzpicture}
        \\
        (a)&(b)
\end{tabular}
\end{center}
\caption{\label{2020-dec-hardy-fig2} (a) Subgraph of the triangle hypergraph
depicted in Fig.~\ref{2020-dec-hardy-fig1} with a faithful orthogonal
representation by decomposable vectors ${\bf b}_2,{\bf b}_2,\ldots,{\bf b}_6$
and arbitrary vectors ${\bf b}_1,{\bf b}_7,{\bf b}_8,{\bf b}_9$. Even though it
turns out that ${\bf b}_1$ and ${\bf b}_7$ must be decomposable, the remaining
vectors ${\bf b}_8$ and ${\bf b}_9$ have to be indecomposable. (b) An explicit
example of a non-faithful orthogonal representation of the triangle hypergraph
with only decomposable vectors resulting in the multiple occurrence of ${\bf s}
\otimes {\bf t}$. }
\end{figure}
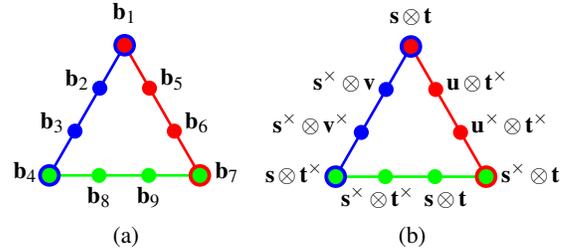

Consider, for the sake of such an example, a ``triangle'' subgraph of the
hypergraph in Fig.~\ref{2020-dec-hardy-fig1}. Suppose we wish to ``dress'' this
hypergraph with a coordinatization involving only decomposable vectors. In
order to show that this task cannot be accomplished we exhibit a faithful
orthogonal representation of the hypergraph depicted in
Fig.~\ref{2020-dec-hardy-fig2}(a). Thereby we merely require
\emph{decomposability} of the vectors ${\bf b}_2,{\bf b}_3,\ldots,{\bf b}_6$
while allowing \emph{arbitrary} vectors ${\bf b}_1,{\bf b}_7,{\bf b}_8,{\bf
b}_9$. Then there exist non-zero vectors ${\bf s}_j,{\bf t}_j\in\mathcal{H}_2$
such that ${\bf b}_j={\bf s}_j\otimes {\bf t}_j$ for all $j=2,3,\ldots,6$.

The plane $(\mathrm{span}\{{\bf b}_1,{\bf b}_7\})^\perp$ contains the two
decomposable vectors ${\bf b}_5$, ${\bf b}_6$ as well as a third
decomposable vector ${\bf b}_4$, all of which are two-by-to linearly
independent. This forces not only $(\mathrm{span}\{{\bf b}_1,{\bf
b}_7\})^\perp$ but also $\mathrm{span}\{{\bf b}_1,{\bf b}_7\}$ to be of type
$(0,0)$ (otherwise there would exist at most two such vectors).
Consequently, there are non-zero vectors ${\bf s}_k,{\bf
t}_k\in\mathcal{H}_2$ such that ${\bf b}_k={\bf s}_k\otimes {\bf t}_k$ for
$k=1,7$. Using Eq.~\eqref{hans-bilinear-det} we arrive at
\begin{equation}\label{hans-bed1}
  ( {\bf b}_1\vert {\bf b}_7 ) = \det( {\bf s}_1, {\bf s}_7) \det( {\bf t}_1, {\bf t}_7)= 0
.
\end{equation}
Furthermore, from  Eqs.~\eqref{hans-inner-rel} and \eqref{hans-inner-det}
we obtain
\begin{equation}\label{hans-bed2}
  \langle {\bf b}_1\vert {\bf b}_7\rangle =
  \langle {\bf s}_1\vert {\bf s}_7\rangle \langle {\bf t}_1\vert {\bf t}_7\rangle =
  \det({\bf s}_1^\times, {\bf s}_7) \det( {\bf t}_1^\times, {\bf t}_7) = 0  .
\end{equation}
Since ${\bf s}_1$ and ${\bf s}_7$ are non-zero, the determinants $\det( {\bf
s}_1, {\bf s}_7)$ and $\det( {\bf s}_1^\times, {\bf s}_7)$ cannot vanish
simultaneously. Likewise, $\det( {\bf t}_1, {\bf t}_7)$ and $\det( {\bf
t}_1^\times, {\bf t}_7)$ are not both zero. Consequently, there are two cases:
(i) either $\det( {\bf s}_1, {\bf s}_7)= \det( {\bf t}_1^\times, {\bf t}_7) = 0
$ and, at the same time, $\det( {\bf t}_1, {\bf t}_7) \neq 0 \neq \det({\bf
s}_1^\times, {\bf s}_7)$, (ii) or, alternatively, $\det({\bf s}_1^\times, {\bf
s}_7) = \det( {\bf t}_1, {\bf t}_7) = 0$ and, at the same time, $\det( {\bf
s}_1, {\bf s}_7) \neq 0 \neq \det( {\bf t}_1^\times, {\bf t}_7)$. Therefore,
either $\det( {\bf s}_1, {\bf s}_7) = 0 \neq \det( {\bf t}_1, {\bf t}_7)$, or,
alternatively, $\det({\bf s}_1^\times, {\bf s}_7) = 0 \neq \det( {\bf
t}_1^\times, {\bf t}_7)$. Hence, up to an irrelevant scaling factor, ${\bf
b}_7={\bf s}_7\otimes{\bf t}_7$ equals one of the following
vectors:
\begin{equation}\label{hans-b4-b7}
    {\bf s}_1\otimes {\bf t}_1^\times,\quad {\bf s}_1^\times \otimes {\bf t}_1 .
\end{equation}

Next, we repeat the previous reasoning in view of ${\bf b}_2, {\bf b}_3,{\bf
b}_7 \in (\mathrm{span}\{{\bf b}_1,{\bf b}_4\})^\perp$. In this way, we regain
the decomposability of ${\bf b}_1$ and ${\bf b}_4$ and arrive at precisely the
same vectors from~\eqref{hans-b4-b7}. So, one of the vectors
from~\eqref{hans-b4-b7} must be proportional to ${\bf b}_4$ while the other
vector needs to be proportional to ${\bf b}_7$. The plane $\mathrm{span}\{{\bf
b}_4,{\bf b}_7\}$ is of type $(1,-1)$ in the real and of type $(1,1)$ in the
complex case, since $({\bf s}_1\otimes {\bf t}_1^\times\vert {\bf s}_1^\times
\otimes {\bf t}_1) = \det({\bf s}_1,{\bf s}_1^\times)\det({\bf t}_1^\times,
{\bf t}_1)\neq 0$. We are therefore in a position to substitute ${\bf s}$ by
${\bf s}_1$, ${\bf t}$ by ${\bf t}_1^\times$, ${\bf u}$ by ${\bf s}_1^\times$
and ${\bf v}$ by ${\bf t}_1$ in \eqref{hans-M-basis}, so that the vectors ${\bf
b}$, ${\bf a}$ appearing in \eqref{hans-Tensorbasis} turn into
\begin{equation}\label{hans-b8+b9}
    {\bf s}_1^\times \otimes {\bf t}_1^\times, \quad (-{\bf s}_1) \otimes (-{\bf t}_1)
    = {\bf s}_1 \otimes {\bf t}_1 = {\bf b}_1 .
\end{equation}
The decomposable vectors from \eqref{hans-b8+b9} constitute a basis of the
plane $\mathrm{span}\{{\bf b}_8,{\bf b}_9\}$. This plane, like its orthogonal
plane $\mathrm{span}\{{\bf b}_4,{\bf b}_7\}$, is of type $(1,-1)$ in the real
and of type $(1,1)$ in the complex case. Thus, up to scaling factors, the
vectors appearing in \eqref{hans-b8+b9} are the only decomposable vectors of
$\mathrm{span}\{{\bf b}_8,{\bf b}_9\}$. Also, we established in
Section~\ref{se:V} that the orthogonality of ${\bf b}_4$ and ${\bf b}_7$ forces
the vectors from \eqref{hans-b8+b9} to be orthogonal. Now, since our orthogonal
representation is faithful, it turns out that ${\bf b}_8$ is not proportional
to ${\bf s}_1 \otimes {\bf t}_1 = {\bf b}_1$, which in turn establishes that
${\bf b}_9$ is not a multiple of ${\bf s}_1^\times \otimes {\bf t}_1^\times$.
The previous statement remains true when interchanging ${\bf b}_8$ and ${\bf
b}_9$. Our final conclusion therefore is that \emph{${\bf b}_8$ and ${\bf b}_9$
have to be indecomposable}.

Fig.~\ref{2020-dec-hardy-fig2}(b) displays an explicit example of a
non-faithful orthogonal representation of a ``triangle'' in terms of
decomposable vectors with just one multiplicity. Thereby, it has to be assumed
that $\{{\bf s},{\bf u}\}$, $\{{\bf s}^\times,{\bf u}\}$, $\{{\bf t},{\bf v}\}$
and $\{{\bf t}^\times,{\bf v}\}$ are bases of $\mathcal{H}_2$ in order to avoid
any further multiplicities. Of course the plane $\mathrm{span}\{{\bf s}^\times
\otimes {\bf t}^\times,{\bf s} \otimes{\bf t}\}$ admits a continuum of
orthogonal bases containing only indecomposable vectors. Replacement of the
given basis $\{{\bf s}^\times \otimes {\bf t}^\times,{\bf s} \otimes{\bf t}\}$
with any such basis yields a faithful orthogonal representation.

\section{Steering (in)decomposability}

If the physical means are restricted to real spaces the existence of ``plain''
planes which contain only non-zero vectors of either one of the two categories
-- factorizable (aka decomposable) and indecomposable -- and the associated
orthogonal planes which are of the same types allows a sort of ``steering''
into such ``plain'' planes. In this way one party controlling the source as
well as the (two elementary) observables spanning the ``original plane'' can,
in a directed manner, signal factorizable or entangled states towards a second
party at the receiving end.

For the sake of an example take two factorizable vectors spanning a type
$(0,0)$ plane, and the associated orthogonal plane which is also of type
$(0,0)$, containing only factorizable vectors. To be more explicit consider a
four-port generalized beam splitter~\cite{rzbb} associated with the output
states corresponding to the vectors $\begin{pmatrix}
1,0,0,0\end{pmatrix}^\intercal$, $\begin{pmatrix}
0,1,0,0\end{pmatrix}^\intercal$, $\begin{pmatrix}
0,0,1,0\end{pmatrix}^\intercal$, and $\begin{pmatrix}
0,0,0,1\end{pmatrix}^\intercal$, respectively. Suppose the first party called
Alice controls the source and the first two ports associated with
$\begin{pmatrix} 1,0,0,0\end{pmatrix}^\intercal$ and $\begin{pmatrix}
0,1,0,0\end{pmatrix}^\intercal$, and the second party called Bob controls the
last two ports associated with $\begin{pmatrix} 0,0,1,0\end{pmatrix}^\intercal$
and $\begin{pmatrix} 0,0,0,1\end{pmatrix}^\intercal$. If Alice makes sure that
she is sending and receiving no other states then she can be sure that Bob, no
matter what he does on ``his side'' of the output ports, will end up with a
factorizable state.

If, on the other hand, only plane types $(1,1)$ are involved
-- say one plane spanned by
$(1/\sqrt{2})\begin{pmatrix} 1,0,0,1\end{pmatrix}^\intercal$ and
$(1/\sqrt{2})\begin{pmatrix} 0,1,1,0\end{pmatrix}^\intercal$
on Alice's state emission and her beam splitter ports,
and
$(1/\sqrt{2})\begin{pmatrix} 1,0,0,-1\end{pmatrix}^\intercal$ and
$(1/\sqrt{2})\begin{pmatrix} 0,1,-1,0\end{pmatrix}^\intercal$
on Bob's ports --
Alice can be sure that Bob, no matter what he does on ``his side'' of the output ports,
will end up with an entangled state.

Note that is suffices for Alice to generate the respective states and observe
her shares of the ports. In that way one can imagine a type of
BB84~\cite{benn-84} protocol which, instead of random shared sequences of bits,
render shared factorizable and entangled states.

We close this investigation into the (in)decomposability of vectors in planes
(aka two-dimensional subspaces) of four-dimensional Hilbert spaces by noting
that their structure exhibits a richness which might not be obvious at first
glance. There exist planes consisting of purely decomposable vectors.
Nevertheless, in general indecomposability and thus physical entanglement and
the encoding of relational properties by quantum states ``prevails'' and occurs
more often than separability associated with well defined individual, separable
states.

\setcounter{MaxMatrixCols}{20}

\begin{acknowledgments}
Karl Svozil acknowledges the support by the Austrian Science Fund (FWF): project I 4579-N and the Czech Science Foundation (GA\v CR): project 20-09869L.

The authors declare no conflict of interest.

The authors kindly acknowledge discussions with Cristian S. Calude and Karol
\.{Z}yczkowski.
\end{acknowledgments}

\section*{Data Availability}
Data sharing is not applicable to this article as no new data were created or analyzed in this study.

\end{document}